\documentclass[pra,aps,preprint,amsfonts,groupedaddress,showpacs,showkeys]{revtex4}
\usepackage{epsfig,tabularx,array,booktabs,calc,multirow,amsmath}
\usepackage{natbib}
\usepackage{bm}

\def\shiftdown#1{#1\llap{\lower.04ex\hbox{#1}}}

\newcommand{\beaa}{\begin{eqnarray*}} 
\newcommand{\enaa}{\end{eqnarray*}}
\newcommand{\bea}{\begin{eqnarray}}
\newcommand{\ena}{\end{eqnarray}}
\newcommand{\be}{\begin{eqnarray}} 
\newcommand{\eq}{\begin{eqnarray}} 
\newcommand{\en}{\end{eqnarray}}

\begin{document}

\title{Radiative decays of the $Y(3940)$, $Z(3930)$ and the $X(4160)$ as dynamically generated resonances} 

\author{
Tanja Branz$^1$, 
Raquel Molina$^2$, 
Eulogio Oset$^2$
\vspace*{1.2\baselineskip}}

\affiliation{$^1$ Institut f\"ur Theoretische Physik,
Universit\"at T\"ubingen,\\ 
Kepler Center for Astro and Particle Physics, \\
Auf der Morgenstelle 14, D--72076 T\"ubingen, Germany
\vspace*{1.2\baselineskip} \\
$^2$ Departamento de F\'{\i}sica Te\'orica and IFIC,
Centro Mixto Universidad de Valencia-CSIC,
Institutos de Investigaci\'on de Paterna, Aptdo. 22085, 46071 Valencia,
 Spain
\\}

\date{\today}

\begin{abstract} 
We study the radiative decay properties of the charmonium-like $X,\,Y$ and $Z$ mesons generated dynamically from vector meson-vector meson interaction in the framework of a unitarized hidden-gauge formalism. In the present work we calculate the one- and two-photon decay widths of the hidden-charm $Y(3940)$, $Z(3930)$ (or X(3915)) and $X(4160)$ mesons in the framework of the vector meson dominance formalism. We obtain good agreement with experiment in case of the two photon width of the $X(3915)$ which we associate with the $2^+$ resonance that we find at 3922 MeV.
\end{abstract}



\maketitle

\newpage

\section{Introduction}

With the discovery of many new unexpected charmonium-like resonances in the last 4-5 years charmonium spectroscopy got back in the focus of interest. In the past the understanding of meson structure was to a large extend based on the constituent quark model. However, it turned out that the meson spectrum is much richer than one might expect from the simple quark model predictions. This fact was first observed in the light meson sector but later on was also found in the heavy quark sector e.g. charmonium spectrum. The majority of the 'new' charmonium-like $X$, $Y$ and $Z$ mesons mainly discovered at the B-factories BELLE and BaBar cannot be easily accommodated in the $q\bar q$ model and are therefore interesting objects for meson structure beyond the constituent quark model. The structure assumptions which have been studied range from threshold effects and hybrids to tetraquarks, meson-meson bound objects and dynamically generated resonances from meson-meson interaction. An overview of the present situation is e.g. given in \cite{Godfrey:2008nc}.

In this context, the coupled channel approach which combines chiral dynamics with a unitarization formalism \cite{Oller:2000ma} turned out to provide a useful tool to determine the mass and width of resonances as well as information on their decay patterns, cross sections etc. In the coupled channel formalism resonances are generated dynamically from meson-meson or meson-baryon interaction and appear as poles in the corresponding scattering amplitude. This approach was first successfully used for the description of pseudoscalar-pseudoscalar meson interaction and pseudoscalars interacting with baryons which therefore suggested a extension to vector meson interaction. Since the chiral Lagrangian only considers interaction with pseudoscalar mesons, the coupled channel approach has to be extended accordingly. The hidden gauge formalism \cite{Bando:1987br,Bando:1984ej,Harada:2003jx,hidden4} provides a consistent method to include vector meson interaction in the above mentioned chiral unitary approach to meson-meson interaction. This method was first applied to $\rho\rho$ meson interaction, leading to dynamically generated $\rho\rho$ resonances around 1.3-1.5 GeV which could be assigned to the $f_0(1370)$ and $f_2(1270)$ \cite{Molina:2008jw}. Subsequently the formalism was extended to SU(4) in order to study the interesting hidden charm resonances around 4 GeV. In \cite{Molina:2009ct} the coupled channel study led to 5 resonances where three of them are good candidates for the $Y(3940)$, $Z(3930)$ and $X(4160)$ mesons discovered by BELLE and BaBar \cite{Abe:2007sya,Abe:2004zs,Uehara:2005qd,Aubert:2007vj}. Two further predictions for resonances have not been seen so far and might be objects for future experimental research.  

A further interesting topic is the study of radiative decay properties which are also a crucial test for hadron structure. In the hidden gauge formalism the electromagnetic interaction is included by using vector meson dominance (VMD) that is the photon couples to the resonance via the $\rho,\omega,\phi$ or $J/\psi$ vector mesons in the respective coupled channels \cite{yamagata,Branz:2009cv,Nagahiro:2008cv}.

In the present paper we concentrate on the hidden-charm resonances around 4 GeV analyzed in \cite{Molina:2009ct} and study the two-photon and photon-vector meson decay properties. The outline is as follows: In the next section we introduce the chiral unitarity approach in combination with the vector meson dominance formalism we use to study radiative decays of dynamically generated resonances. In section \ref{sec:res} we present our results for the $\gamma\gamma$ and $V\gamma$ radiative decays of the charmonium-like $Y(3940)$, $Z(3930)$ and $X(4160)$ mesons. We compare the decay properties with available data and other theoretical approaches provided they exist. Finally we summarize our work in section \ref{sec:sum}.

\section{Coupled channel approach}

\begin{figure}[htbp]
\includegraphics[scale=0.6]{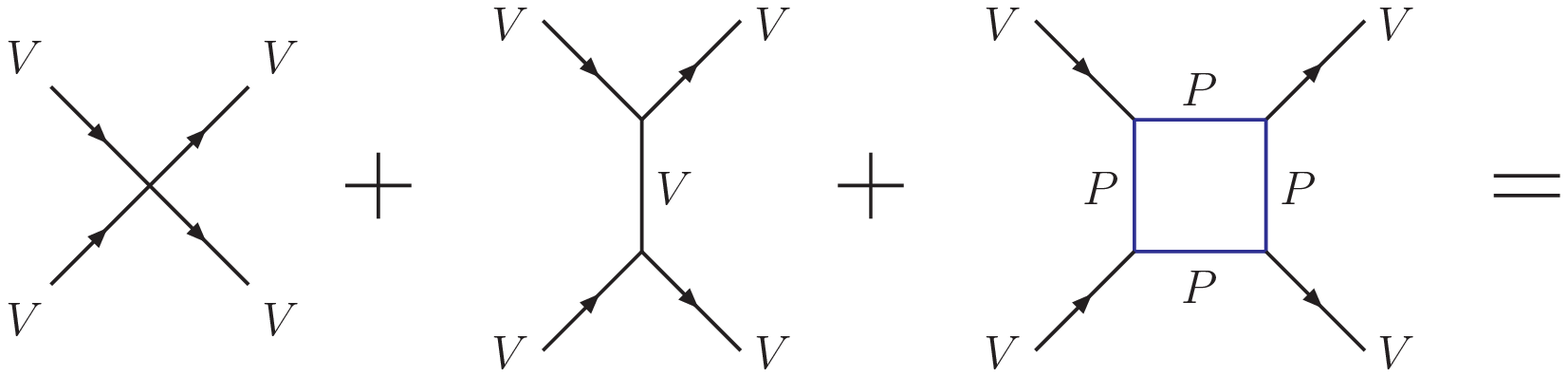}
\caption{$VV$ interaction diagrams.}
\label{fig1} 
\end{figure}

\begin{figure}[htbp]
\includegraphics[scale=0.6]{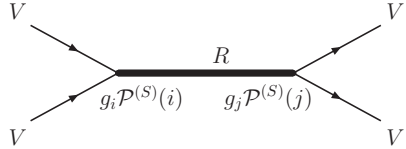}
\caption{Dynamically generated resonance from Bethe Salpeter equation}
\label{fig2} 
\end{figure}

In the coupled channel model resonances are generated by meson-meson interaction under consideration of unitarization \cite{Oller:2000ma}. The scattering amplitude $T$ is set up by means of the Bethe-Salpeter equation 
\eq
T=(1-VG)^{-1}V\,,\label{eq:BS}
\en
where the kernel $V$ is provided by the interaction between different meson channels (see Fig. \ref{fig1}) and $G$ is a diagonal matrix of the respective meson loops. The meson loop integrals are UV divergent and are regularized by either cutoff or dimensional regularization. For further detail we refer to \cite{Molina:2008jw,Molina:2009ct}. In the present case we consider coupled channels between vector mesons, where the interaction is given by the hidden gauge Lagrangian. We include four vector contact interaction given by
\eq
{\cal L}_{III}^{(c)}=\frac{g^2}{2}\big<V_\mu V_\nu V^\mu V^\nu-V_\nu V_\mu V^\mu V^\nu\big>\,,\label{eq:L1}
\en
which is diagrammatically represented by Fig. \ref{fig1} a). $V^\mu$ is the SU(4) matrix containing the interacting vector mesons. Further on we consider vector meson exchange terms (see Fig. \ref{fig1} b)) which are based on the three vector meson interaction Lagrangian
\eq
{\cal L}_{III}^{(3V)}=ig\big<(\partial_\mu V_\nu-\partial_\nu V_\mu)V^\mu V^\nu\big>\label{eq:L2}\,.
\en
These two mechanisms are responsible for the generation of resonances as indicated in Fig. \ref{fig2} provided that the interaction is sufficiently strong.

While four vector contact terms and vector-exchange terms determine the mass and part of the width of the resonances, the two pseudoscalar decay modes are only relevant for the generation of the widths of the resonances. The corresponding box diagrams with intermediate pseudoscalars in Fig. \ref{fig1} c) are based on the interaction Lagrangian
\eq
{\cal L}_{V\Phi\Phi}=-ig\big<V^\mu\big[\Phi,\partial_\mu \Phi\big]\big>\label{eq:L3}\,.
\en
The transition amplitude $T$ between the initial and final coupled channels via resonance $R$ from the Bethe Salpeter equation in Eq. (\ref{eq:BS}) can be approximated close to a pole by
\be
T_{ij}^{(S)}=g_i{\cal P}^{(S)}(i)\frac{1}{s-M_R^2+iM_R\Gamma_R}g_j{\cal P}^{(S)}(j)\,,\label{eq:T}
\en
where
\be
{\cal P}^{(0)}&=&\frac{1}{\sqrt3}\epsilon_i(1)\epsilon_i(2)\,,\label{eq:p1}\\
{\cal P}^{(1)}&=&\frac12\big[\epsilon_i(1)\epsilon_j(2)-\epsilon_j(1)\epsilon_i(2)\big]\,,\label{eq:p2}\\
{\cal P}^{(2)}&=&\frac12\big[\epsilon_i(1)\epsilon_j(2)+\epsilon_j(1)\epsilon_i(2)\big]-\frac13\epsilon_m(1)\epsilon_m(2)\delta_{ij}\,,\label{eq:p3}
\en
denote the  spin projection operators of the respective coupled channels. Here $\epsilon(1)$ and $\epsilon(2)$ are the polarization vectors of particle 1 and 2, respectively. The indices $i,j$ and $m$ run over the spatial coordinates, i.e. $i,j,m=1,2,3$.

The couplings of the resonance $R$ to each particular $VV$ channel can be extracted from the residues of the $T$-matrix at the poles because of the approximation in Eq. (\ref{eq:T}). The coupling strength to each particular channel gives an idea of the importance of its contribution to the resonance. In case of the $Y(3940)$ and $Z(3930)$ the $D^\ast\bar D^\ast$ coupling is dominant, which implies, e.g., that the $Y(3940)$ is dominantly a $D^\ast\bar D^\ast$ bound state, in agreement with the $D^\ast \bar D^\ast$ molecular interpretations in \cite{Liu:2009ei,Branz:2009yt} while the $X(4160)$ is mainly $D_s^\ast\bar D_s^\ast$.

\begin{figure}[htbp]
\begin{center}
\includegraphics[scale=0.6]{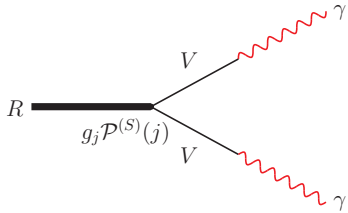}
\end{center}
\caption{Photon coupling via VMD}
\label{fig3} 
\end{figure}

We study the radiative decays of dynamically generated resonances by coupling the photon via intermediate vector mesons as depicted in Fig. \ref{fig3}. 
The resulting amplitude of the $V\gamma$ transition is provided by the vector meson dominance formalism which leads to
\eq
t_{V\gamma}={\cal C}_{V\gamma}\frac eg M_V^2\epsilon_\mu(V)\epsilon^\mu(\gamma)\text{ with }
{\cal C}_{V\gamma}=\left\{\begin{array}{cl}\frac{1}{\sqrt2}&\text{ for }\rho^0\\
\frac{1}{3\sqrt2}&\text{ for }\omega\\
-\frac13&\text{ for }\phi \\
\frac23&\text{ for }J/\psi\,,\end{array}\right.
\en
where we fix $g$ by $g=\frac{m_\rho}{2f_\pi}$ with $f_\pi=93$ MeV. In case of the charmonium, i.e. $J/\psi$, we consider $SU(4)$ breaking effects by using $g\equiv g_{\eta_c}=M_{J/\psi}/(2f_{\eta_c})$, where $f_{\eta_c}=420/\sqrt2$ MeV is taken from \cite{Sun:2008ew}.

The amplitude for the one- and two-photon decays are therefore given by
\eq
T^{(R)}(\gamma\gamma)&\propto&\sum\limits_{V_1,V_2}g_{V_1V_2}^{(R)}{\cal P}_{V_1V_2}^{(S)}\Big(\frac{1}{-M_{V_1}^2}\Big)t_{V_1\gamma}\Big(\frac{1}{-M_{V_2}^2}\Big)t_{V_2\gamma}\cdot F_I\,,\label{eq:11}\\
T^{(R)}(V_1\gamma)&\propto&\sum\limits_{V_2}g_{V_1V_2}^{(R)}{\cal P}_{V_1V_2}^{(S)}\Big(\frac{1}{-M_{V_2}^2}\Big)t_{V_2\gamma}\cdot F_I\,,\label{eq:12}
\en
where  $g_{V_1V_2}$ is the coupling of the resonance $R$ to the $V_1V_2$ channel (see Tables \ref{tab1} and \ref{tab2}) and $F_I$ represents the respective isospin Clebsch Gordan coefficients of the $V_1V_2$ component for a certain isospin state. The isospin states are given by
\eq
\big|D^\ast\bar D^\ast,I=0,I_3=0\big>&=&\frac{1}{\sqrt2}\big(\big|D^{\ast\,+} D^{\ast\,-}\big>+\big|D^{\ast\,0}\bar D^{\ast\,0}\big>\big)\,,\\
\big|D^\ast\bar D^\ast,I=1,I_3=0\big>&=&\frac{1}{\sqrt2}\big(\big|D^{\ast\,+} D^{\ast\,-}\big>-\big|D^{\ast\,0}\bar D^{\ast\,0}\big>\big)\,,\\
\big|\rho\rho,I=0,I_3=0\big>&=&-\frac{1}{\sqrt3}\big(\big|\rho^+\rho^-\big>+\big|\rho^-\rho^+\big>+\big|\rho^0\rho^0\big>\big)\,,\\
\big|\rho\rho,I=1,I_3=0\big>&=&-\frac{1}{\sqrt2}\big(\big|\rho^+\rho^-\big>-\big|\rho^-\rho^+\big>\big)\,,
\en
when using the phase convention $(-D^{\ast\,0},D^{\ast\,+})$, $(D^{\ast\,-},\bar D^{\ast\,0})$ and $(-\rho^+,\rho^0,\rho^-)$ for the respective isospin doublets and triplet. After summing over the intermediate vector polarizations in Eqs. (\ref{eq:11}) and (\ref{eq:12}),
the amplitudes for the $R\to\gamma\gamma$ and $R\to V\gamma$ decays are finally given by
\eq
T^{( R)}_{V_1\gamma}&=&\frac{e}{g}\sum\limits_{V_2=\rho^0,\omega,\phi,J/\psi}g_{V_1V_2}^{(R)}{\cal P}_{V_1\gamma}^{(S)}{\cal C}_{V_2\gamma}\times F_I\times F_{V\gamma} \,,\\
T^{(R)}_{\gamma\gamma}&=&\frac{e^2}{g^2}\sum\limits_{V_1,V_2=\rho^0,\omega,\phi,J/\psi}g_{V_1V_2}^{(R)}{\cal P}_{\gamma\gamma}^{(S)}{\cal C}_{V_1\gamma}{\cal C}_{V_2\gamma}\times F_I\times F_{{\gamma\gamma}}\,,
\en
where ${\cal P}_{V_1\gamma}^{(S)}$ and ${\cal P}_{\gamma\gamma}^{(S)}$ are the spin projection operators of Eqs. (\ref{eq:p1}), (\ref{eq:p2}) and (\ref{eq:p3}) with the polarizations of $V\gamma$ in the first case and of $\gamma\gamma$ in the second case.
Due to the use of the unitarity renormalization with an extra factor $\frac{1}{\sqrt2}$ in case of identical particles we need to correct this factor when calculating observables. The unitarity normalization and the symmetry factors are combined in $F_{{\gamma\gamma}}$ and $F_{V\gamma}$ with
\eq
F_{\gamma\gamma}&=&\left\{\begin{array}{cl}\sqrt2&\text{for a pair of identical particles, e.g $\rho^0\rho^0$}\\2&\text{for a pair of different particles, e.g $\rho^0\omega$}\,,\end{array}\right.\\
F_{V\gamma}&=&\left\{\begin{array}{cl}\sqrt2&\text{for a pair of identical particles, e.g $\rho^0\rho^0$}\\1&\text{for a pair of different particles, e.g $\rho^0\omega$}\,.\end{array}\right.
\en
The radiative decay widths $\Gamma_{\gamma\gamma}$ and $\Gamma_{V\gamma}$ can be easily calculated from the transition amplitudes $T$ by the relations
\eq
\Gamma_{\gamma\gamma}&=&\frac{1}{2S+1}\frac{1}{16\pi M_R}\frac12\cdot\sum\limits_{spins}\big|T_{\gamma\gamma}^{(R)}\big|^2\,,\\
\Gamma_{V\gamma}&=&\frac{1}{2S+1}\frac{1}{8\pi M_R}\frac{\big|\vec{p}_\gamma|}{M_R}\cdot\sum\limits_{spins}\big|T_{V\gamma}^{(R)}\big|^2\,,
\en
where the summation over all spin states contributes the factors
\eq
\sum\limits_{spins}{\cal P}_{\gamma\gamma}^{(S)}{\cal P}_{\gamma\gamma}^{\ast(S)}&=&\left\{\begin{array}{ll}
\frac23&S=0\\
1&S=1\\
\frac73&S=2\,,
\end {array}\right.\\
\sum\limits_{spins}{\cal P}_{V\gamma}^{(S)}{\cal P}_{V\gamma}^{\ast(S)}&=&\left\{\begin{array}{ll}
\frac23&S=0\\
2&S=1\\
\frac{10}{3}&S=2\,.
\end {array}\right.
\en

\section{results}\label{sec:res}
In order to compute the decay widths of the $Y(3940)$, $Z(3930)$, $X(4160)$ and the so far not observed $'Y_p(3912)'$ we take the couplings between resonances and the $VV$ channels from \cite{Molina:2009ct}. The couplings for isosinglet resonances are given in Tab. \ref{tab1} while the couplings of the $I=1$ state are indicated in Tab. \ref{tab2}. In the case of the state with quantum numbers $I^G(J^{PC})=0^-\,(1^{+-})$ predicted in \cite{Molina:2009ct} all couplings to vector mesons with hidden flavor are zero due to $C$-parity violation. Therefore radiative decays via VMD are forbidden in this case. 
\begin{table}[htbp]
\begin{center}
     \renewcommand{\arraystretch}{1.6}
     \setlength{\tabcolsep}{0.6cm}
\begin{tabular}{cccc}\hline\hline
Channel&\multicolumn{3}{c}{pole positions and $I^G[J^{PC}]$}\\\hline
&$3943 + i 7.4$, $\;0^+[0^{++}]$&$3922+i 26$, $\;0^+[2^{++}]$&$4169+i 66$, $\;0^+[2^{++}]$\\\hline\hline
$\rho\rho$&$-22 + i 47$&$-75 +i 37$&$70+ i20$\\\hline
$\omega\omega$&$1348+ i 234 $&$1558 + i 1821$&$3 -i 2441$\\\hline
$\phi\phi$&$-1000 -i 150$&$-904 - i1783 $&$1257+ i 2866 $\\\hline
$J/\psi J/\psi$&$417+ i 64$&$1783 +i 197$&$2681+ i 940$\\\hline
$\omega\phi$&$-215 - i107$&$91 -i 784$&$1012+ i 1522$\\\hline
$\omega J/\psi$&$-1429 - i 216$&$-2558 - i2289$&$-866 + i 2752 $\\\hline
$\phi J/\psi$&$889+ i 196 $&$918+ i2921 $&$-2617 - i5151 $\\\hline\hline
\end{tabular}
\end{center}
\caption{Couplings $g_{i}$ in units of MeV for the resonances with $I=0$.}
\label{tab1}
\end{table} 
\begin{table}[htbp]
\begin{center}
 \renewcommand{\arraystretch}{1.6}
     \setlength{\tabcolsep}{0.4cm}
\begin{tabular}{cccccc}
\hline
pole position& $I^G[J^{PC}]$&$\rho\rho$&$\rho\omega$&$\rho J/\psi$&$\rho\phi$\\
\hline
\hline
$3919+ i74$& $1^-[2^{++}]$&$0$&$-1150 -i 3470 $&$2105+ i5978$&$-1067 -i 2514$\\
\hline
\end{tabular}
\end{center}
\caption{Couplings $g_{i}$ in units of MeV for $I=1$, $J=2$.}
\label{tab2}
\end{table} 

The results for the radiative decay widths are summarized in Tab. \ref{tab3}. The decay widths of the $Y(3940)$ are in general smaller compared to the other resonances. A further common feature is that the $\rho\gamma$ and $\gamma\gamma$ decay modes are suppressed in comparison to the $\omega\gamma$ and $\phi\gamma$ decays except for the predicted $Y_p(3912)$ resonance, which shows a rather strong coupling to the $\rho\gamma$ decay channel. 
\begin{table}[htbp]
 \begin{center}
      \renewcommand{\arraystretch}{1.6}
     \setlength{\tabcolsep}{0.3cm}
     \centering
\begin{tabular}{cccccccc}
\hline\hline
pole [MeV]&$I^G\,J^{PC}$&meson&$\Gamma_{\rho \gamma}$[keV]&$\Gamma_{\omega \gamma}$[keV]&$\Gamma_{\phi \gamma}$[keV]&$\Gamma_{J/\psi \gamma}$[keV]&$\Gamma_{\gamma \gamma}$[keV]\\\hline\hline
$(3943, +i 7.4)$&$0^+\,(0^{++})$&$Y(3940)$&$0.015$&$0.989$&$13.629$&$0.722$&$0.013$\\
\hline
$(3922,+ i 26)$&$0^+\,(2^{++})$&$Z(3930)$&$0.040$&$15.155$&$95.647$&$13.952$&$0.083$\\
\hline
$(4169,+ i 66)$&$0^+\,(2^{++})$&$X(4160)$&$0.029 $&$10.659$&$268.854$&$125.529$&$0.363$\\
\hline
$(3919,+ i 74)$&$1^-\,(2^{++})$&$'Y_p(3912)'$&$201.458$&$114.561$&$62.091$&$135.479$&$0.774$\\
\hline\hline
\end{tabular}
\end{center}
\caption{Pole positions and radiative decay widths.}\label{tab3}
\end{table} 
The relatively small two-photon decay width $\Gamma(Y(3940)\to \gamma\gamma)=0.013$ keV underestimates the corresponding result in the hadronic molecule interpretation of $\Gamma(Y(3940)\to \gamma\gamma)=0.33$ keV for $J^{PC}=0^{++}$ in \cite{Branz:2009yt} by more than one order of magnitude. However, the coupled channel analysis also regards the $D^\ast\bar D^\ast$ component as dominant. Yet, we observe that in terms of $VV$ loops to which the two photons couple, there is a strong cancellation between the contribution of $D^*\bar{D}^*$ loops and $\rho\rho$ loops, the latter ones not considered in \cite{Branz:2009yt}.  Even when restricting to the $D^*\bar{D}^*$ loop, substantial differences from the results of \cite{Branz:2009yt} remain. As a consequence of the cancellations  found in the $\gamma\gamma$ rate, the uncertainties can be regarded to be large, we deal with about a factor two or three.
The differences with the results of \cite{Branz:2009yt} might be an indication of probably inaccurate approximations in our scheme (we will improve this result below). One of the approximations done in \cite{Molina:2009ct} is to neglect the three momenta of the particles with respect to the vector mass, $\vert\vec{p}\vert/M_{D^*}\simeq 0$. This is indeed the case for massive states but not if we deal with two photons in the final state since $\vert\vec{p}_\gamma\vert\simeq M_{D^*}$. Hence, improvements on the present results are necessary. For this purpose we reanalyze the couplings of \cite{Molina:2009ct} by considering the final momenta of the light vectors which finally couple to the photons.
We therefore modify the corresponding expressions in  \cite{Molina:2009ct} in the following way, suited for the use of the Coulomb gauge for the photons.
\begin{itemize}
\item[1)] For $J=0$, we use:
\begin{eqnarray}
(k_{1}^0+k_{3}^0)(k_{2}^0+k_{4}^0)\to (k_{1}^0+k_{3}^0)(k_{2}^0+k_{4}^0) +5\vec{k}^2_3+\frac{\vec{k}^4_3}{M_{\mathrm{exc}}^2}
\end{eqnarray}
\item[2)] For $J=2$, we change:
\begin{eqnarray}
(k_{1}^0+k_{3}^0)(k_{2}^0+k_{4}^0)\to \sqrt{a^2+\frac{32}{5}\vec{k}_3^2}
\label{eq:1}
\end{eqnarray}
with
\begin{eqnarray}
a=(k_{1}^0+k_{3}^0)(k_{2}^0+k_{4}^0) +\vec{k}^2_3+\frac{\vec{k}^4_3}{M_{\mathrm{exc}}^2}\ ,
\label{eq:2}
\end{eqnarray}
where $\vec{k}_3$ is the photon momentum and $M_{\mathrm{exc}}$ is the mass of the exchanged vector meson, $M_{D^*}$ or $M_{D^*_s}$.
\end{itemize}
With this prescription we run the coupled channels Bethe-Salpeter equations and obtain new couplings of the resonances to pairs of vectors, which, via the VMD, provide the appropriate couplings of the resonances to the photons. The couplings to $D^*\bar{D}^*$ and $D^*_s\bar{D}^*_s$ are practically unchanged, as well as the masses of the resonances. However we obtain new effective couplings to the light vectors which characterize the $\gamma\gamma$ decay rates. The new results are given in Table \ref{tab:nw}.
\begin{table}[htbp]
 \begin{center}
      \renewcommand{\arraystretch}{1.6}
     \setlength{\tabcolsep}{0.3cm}
     \centering
\begin{tabular}{cccc}
\hline\hline
pole [MeV]&$I^G\,J^{PC}$&meson&$\Gamma^{\mathrm{new}}_{\gamma \gamma}$[keV]\\\hline\hline
$(3943, +i 7.4)$&$0^+\,(0^{++})$&$Y(3940)$&$0.085$\\
\hline
$(3922,+ i 26)$&$0^+\,(2^{++})$&$Z(3930)$&$0.074$\\
\hline
$(4169,+ i 66)$&$0^+\,(2^{++})$&$X(4160)$&$0.54$\\
\hline
$(3919,+ i 74)$&$1^-\,(2^{++})$&$'Y_p(3912)'$&$1.11$\\
\hline\hline
\end{tabular}
\end{center}
\caption{New values of the $\Gamma^{\mathrm{new}}_{\gamma \gamma}$ after running the coupled channel Bethe-Salpeter equations with the modifications of Eqs. (\ref{eq:1}) and (\ref{eq:2}).}
\label{tab:nw}
\end{table} 
Due to the inclusion of the three-momenta of the photons the decay widths undergo changes of
 $30\%$ in all cases, except for the Y(3940) state, where the abnormally small former width becomes by a factor of six larger. Our new result lies much closer to the prediction of  \cite{Branz:2009yt} based on a $D^*\bar{D}^*$ molecular structure interpretation of the $Y(3940)$. The two photon width obtained within the coupled channel formalism is still by a factor three smaller than the result of the effective Lagrangian approach of \cite{Branz:2009yt} but this discrepancy lies within the  theoretical uncertainties  of this small rate. The inclusion of the photon three-momenta also leads to an increase of the $\gamma V$  decay widths quoted in Table \ref{tab3} by about a factor of two in case of the Y(3940). However this deviation can be absorbed in the uncertainty of our results.

For the two-photon width of the X(4160) we obtain $\Gamma^{\mathrm{new}}_{\gamma\gamma}=0.54$ keV. In the present coupled channel approach the X(4160) is found to be dominantly a  $D^*_s\bar{D}^*_s$ state. This is the same underlying structure as the $D_s^\ast \bar D_s^\ast$ bound state studied in~\cite{Branz:2009yt}. In reference~\cite{Branz:2009yt} the $D_s^\ast \bar D_s^\ast$ molecular state was associated with the narrow Y(4140) discovered by the CDF \cite{CDF} because it was possible to explain the sizable observed $J/\psi \phi$ decay width of this state. Our association to the broader X(4160) is suggested by the large total theoretical width which was not evaluated in \cite{Branz:2009yt}.
Since the nature of the resonances is the same in both approaches it is less surprising that our result for $\Gamma_{\gamma\gamma}=0.54$ keV agrees with the $\Gamma_{\gamma\gamma}=0.5$ keV evaluated in \cite{Branz:2009yt}.

Experimental observations concerning the radiative decays are rare. However, the BELLE Collaboration searched for charmonium-like resonances in the $\gamma\gamma\to \omega J/\psi$ process \cite{:2009tx} which resulted in an enhancement of the cross section around M = $3915\pm3\pm2$ MeV. The peak was associated to a resonance denoted by $X(3915)$. But it is thought that it could be the $Y(3940)$ resonance, or even the Z(3930) which we have associated to our $J^P=2^+$ resonance at $3922$ MeV. In \cite{:2009tx}, the X(3915) has unknown spin and parity, but $0^+$ or $2^+$ are preferred. In the following we compare the ratios
\begin{eqnarray}\label{eq1}
\Gamma_{\gamma\gamma}(X(3915)){\cal B}(X(3915)\to \omega J/\psi)=\left\{\begin{array}{cl}(61\pm17\pm8) \text{ eV}&\text{ for }J^P=0^+\\
(18\pm5\pm2) \text{ eV}&\text{ for }J^P=2^+\\
\end{array}\right.
\end{eqnarray}
quoted in \cite{:2009tx}  with the results of the present approach. Let us evaluate Eq. (\ref{eq1}) for the two theoretical states $0^+$ at $3943$ MeV and $2^+$ at $3922$ MeV in Table \ref{tab:nw}. By using the simple formula for the decay width to $\omega J/\psi$:
\eq\label{eqwidth}
\Gamma_{\omega J/\psi}=\frac{1}{8\pi}\frac{kg_{\omega J/\psi}^2}{M_R^2}\ ,
\en
with $k$ the momentum of the final meson, and taking the couplings $g_{\omega J/\psi}$ from Table~\ref{tab1}, we obtain
\eq
\Gamma_{\omega J/\psi(0^+,3943)}&=&1.52\text{ MeV}\nonumber\\
\Gamma_{\omega J/\psi(2^+,3922)}&=&8.66\text{ MeV}\,.
\en
Together with the two photon decay widths of Table \ref{tab:nw} we find
\begin{eqnarray}
\Gamma_{\gamma\gamma}{\cal B}((0^+, 3943)\to \omega J/\psi)&=&7.6\text{ eV}\nonumber\\
\Gamma_{\gamma\gamma}{\cal B}((2^+, 3922)\to \omega J/\psi)&=&11.8 \text{ eV}\ .
\label{eq2}
\end{eqnarray}
Comparing these results with the experimental predictions in Eq. (\ref{eq1}) and considering $20\%$ uncertainties in the $\gamma\gamma$ rates for the ($2^+$, $3922$) state, the results of Eq. (\ref{eq2}) are compatible with the assumption that the X(3915) is the resonance ($2^+$, $3922$), considered as the Z(3930) in Table \ref{tab3}.

On the other hand, if we assume that the X(3915) corresponds to our ($0^+$, $3943$) resonance, the discrepancies are very large, even if we accept a factor three uncertainty in our result for $\Gamma_{\gamma\gamma}$. Our study, thus, favors the association of our ($2^+$, $3922$) resonance to the X(3915) of \cite{:2009tx}.

\section{conclusions}\label{sec:sum}
We studied the two-photon and photon-vector meson decay properties of dynamically generated resonances from vector-vector coupled channels in a unitarity hidden gauge formalism. In the present work the focus was set on the hidden-charm mesons around 4 GeV analyzed in \cite{Molina:2009ct}. According to their masses and widths three of them are good candidates for the $Y(3940)$, $Z(3930)$ and $X(4160)$ mesons discovered by BELLE and BaBar. Further on the $Z(3930)$ possibly corresponds to the recently observed $X(3915)$. The two-photon decay width of the $Y(3940)$ in the hidden gauge formalism, assumed to be $0^+$, is more uncertain due to large cancellations, and could be compatible with the larger width of the prediction obtained in the pure $D^\ast \bar D^\ast$ molecule interpretation in \cite{Branz:2009yt}.
In the case of the X(4160), which we assume to be the $2^+$ state at $4169$ MeV, the $\gamma\gamma$ decay width agrees with the results of \cite{Branz:2009yt}, for a $D^*_s\bar{D}^*_s$ molecule with a chosen mass of $4140$ MeV.

The information on the $\Gamma_{\gamma\gamma}$ decay rate of the X(3915) favors the association of this resonance to the ($2^+$, $3922$) resonance that we obtain. Here, the quantum numbers $0^+$ are clearly disfavored.

Unfortunately there is not much data on radiative decays of the $X, Y$ and $Z$ mesons. The large variety of results obtained by us concerning the different decays and different resonances indicates that these measurements are very useful to  shed light on the structure of these resonances. In particular the vector meson-photon decay modes could be addressed in future facilities like PANDA and BESIII, and undoubtedly these measurements would be very valuable to help to disentangle different aspects of the nature of these charmonium like states.

\begin{acknowledgments}
T.B. acknowledges support from the DFG under Contract No. GRK683. This research is partly supported by the DGICYT contract number FIS2006-03438 and the Generalitat Valenciana in the Prometeo Program. We acknowledge the support of the European Community-Research Infrastructure Integrating Activity "Study of Strongly Interacting Matter" (acronym HadronPhysics2, Grant Agreement n. 227431) under the Seventh Framework Progamme of EU.
\end{acknowledgments}


\newpage
\begin{thebibliography}{99} 
\bibitem{Godfrey:2008nc}
  S.~Godfrey and S.~L.~Olsen,
  Ann.\ Rev.\ Nucl.\ Part.\ Sci.\  {\bf 58}, 51 (2008)

\bibitem{Oller:2000ma}
  J.~A.~Oller, E.~Oset and A.~Ramos,
  Prog.\ Part.\ Nucl.\ Phys.\  {\bf 45} (2000) 157
\bibitem{Bando:1987br}
  M.~Bando, T.~Kugo and K.~Yamawaki,
  Phys.\ Rept.\  {\bf 164} (1988) 217.

\bibitem{Bando:1984ej}
  M.~Bando, T.~Kugo, S.~Uehara, K.~Yamawaki and T.~Yanagida,
  Phys.\ Rev.\ Lett.\  {\bf 54} (1985) 1215.

\bibitem{Harada:2003jx}
  M.~Harada and K.~Yamawaki,
  Phys.\ Rept.\  {\bf 381}, 1 (2003)
\bibitem{hidden4}
U.~G.~Meissner,
Phys.\ Rept.\  {\bf 161}, 213 (1988).

\bibitem{Molina:2008jw}
  R.~Molina, D.~Nicmorus and E.~Oset,
  Phys.\ Rev.\  D {\bf 78}, 114018 (2008)


\bibitem{Molina:2009ct}
  R.~Molina and E.~Oset,
  Phys.\ Rev.\  D {\bf 80}, 114013 (2009)

\bibitem{Abe:2007sya}
  P.~Pakhlov {\it et al.}  [Belle Collaboration],
  Phys.\ Rev.\ Lett.\  {\bf 100}, 202001 (2008)
\bibitem{Uehara:2005qd}
  S.~Uehara {\it et al.}  [Belle Collaboration],
  Phys.\ Rev.\ Lett.\  {\bf 96} (2006) 082003

\bibitem{Abe:2004zs}
  K.~Abe {\it et al.}  [Belle Collaboration],
  Phys.\ Rev.\ Lett.\  {\bf 94}, 182002 (2005)


\bibitem{Aubert:2007vj}
  B.~Aubert {\it et al.}  [BaBar Collaboration],
  Phys.\ Rev.\ Lett.\  {\bf 101}, 082001 (2008)
\bibitem{yamagata}
  H.~Nagahiro, J.~Yamagata-Sekihara, E.~Oset and S.~Hirenzaki,
  Phys.\ Rev.\  D {\bf 79}, 114023 (2009)
  [arXiv:0809.3717 [hep-ph]].


\bibitem{Branz:2009cv}
  T.~Branz, L.~S.~Geng and E.~Oset,
  Phys.\ Rev.\  D {\bf 81}, 054037 (2010)


\bibitem{Nagahiro:2008cv}
  H.~Nagahiro, L.~Roca, A.~Hosaka and E.~Oset,
  [arXiv:0809.0943 [hep-ph]].



  
\bibitem{Liu:2009ei}
  X.~Liu and S.~L.~Zhu,
  Phys.\ Rev.\  D {\bf 80}, 017502 (2009)
\bibitem{Branz:2009yt}
  T.~Branz, T.~Gutsche and V.~E.~Lyubovitskij,
  Phys.\ Rev.\  D {\bf 80}, 054019 (2009)
 
\bibitem{Sun:2008ew}
  J.~F.~Sun, D.~S.~Du and Y.~L.~Yang,
  Eur.\ Phys.\ J.\  C {\bf 60}, 107 (2009)


  \bibitem{CDF}
  T.~Aaltonen et al. (The CDF collaboration)
 [arXiv:0903.2229 [hep-ph]]

\bibitem{:2009tx}
  S.~Uehara {\it et al.}  [Belle Collaboration],
  Phys.\ Rev.\ Lett.\  {\bf 104}, 092001 (2010)
\end{thebibliography}
\end{document}